\documentstyle[12pt]{article}
\catcode`\@=11

\@addtoreset{equation}{section}

\def\@normalsize{\@setsize\normalsize{15pt}\xiipt\@xiipt
\abovedisplayskip 14pt plus3pt minus3pt%
\belowdisplayskip \abovedisplayskip
\abovedisplayshortskip  \z@ plus3pt%
\belowdisplayshortskip  7pt plus3.5pt minus0pt}
\def\small{\@setsize\small{13.6pt}\xipt\@xipt
\abovedisplayskip 13pt plus3pt minus3pt%
\belowdisplayskip \abovedisplayskip
\abovedisplayshortskip  \z@ plus3pt%
\belowdisplayshortskip  7pt plus3.5pt minus0pt
\def\@listi{\parsep 4.5pt plus 2pt minus 1pt
            \itemsep \parsep
            \topsep 9pt plus 3pt minus 3pt}}

\def\underline#1{\relax\ifmmode\@@underline#1\else
        $\@@underline{\hbox{#1}}$\relax\fi}
\@twosidetrue
\relax

\catcode`@=12

\catcode`\@=11

\def\section{\@startsection{section}{1}{\z@}{3.5ex plus 1ex minus
   .2ex}{2.3ex plus .2ex}{\large\bf}}

\def\ps@headings{\def\@oddfoot{}\def\@evenfoot{}
\def\@oddhead{\hbox{}\hfill
        \makebox[.5\textwidth]{\raggedright\ignorespaces --\thepage{}--
        \hfill }}
\def\@evenhead{\@oddhead}
\def\subsectionmark##1{\markboth{##1}{}}
}

\ps@headings

\catcode`\@=12

\relax

\def\figcap{\section*{Figure Captions\markboth
        {FIGURECAPTIONS}{FIGURECAPTIONS}}\list
        {Fig. \arabic{enumi}:\hfill}{\settowidth\labelwidth{Fig. 999:}
        \leftmargin\labelwidth
        \advance\leftmargin\labelsep\usecounter{enumi}}}
 \relax
\def\tablecap{\section*{Table Captions\markboth
        {TABLECAPTIONS}{TABLECAPTIONS}}\list
        {Table \arabic{enumi}:\hfill}{\settowidth\labelwidth{Table 999:}
        \leftmargin\labelwidth
        \advance\leftmargin\labelsep\usecounter{enumi}}}
 \relax
\def\reflist{\section*{References\markboth
        {REFLIST}{REFLIST}}\list
        {[\arabic{enumi}]\hfill}{\settowidth\labelwidth{[999]}
        \leftmargin\labelwidth
        \advance\leftmargin\labelsep\usecounter{enumi}}}
 \relax

\catcode`\@=11

\def\ps@headings{\def\@oddfoot{}\def\@evenfoot{}
\def\@oddhead{\hbox{}\hfill
        \makebox[.5\textwidth]{\raggedright\ignorespaces --\thepage{}--
        \hfill }}
\def\@evenhead{\@oddhead}
\def\subsectionmark##1{\markboth{##1}{}}
}

\ps@headings

\relax

\def\firstpage#1#2#3#4#5#6{
\begin{document}
\begin{titlepage}
\nopagebreak
\title{\begin{flushright}
        \vspace*{-1.5in}
        {\normalsize LBL--36744 -- UCB-95/03\\[-3mm]\normalsize
NUB--#1\\[-3mm]LPTHE-Orsay 95/14 \\ [-3mm]
#2\\[6mm]} \end{flushright}
\vfill
{\large \bf #3}}
\author{\large #4 \\[1cm] #5}
\maketitle
\vfill
\nopagebreak
\begin{abstract}
{\noindent #6}
\end{abstract}
\vfill
\begin{flushleft}
\rule{16.1cm}{0.2mm}\\[-3mm]
$^{\star}${\small Research supported in part by the Director,\vspace{-4mm}
Office of High Energy and Nuclear Physics, Division of High Energy
Physics of \vspace{-4mm} the U.S. Department of Energy under contract
DE-AC03-76SF00098 and in part by the National Science\vspace{-4mm}
Foundation under grants PHY--90--21139 and
PHY--93--06906.}\\
$^{\dagger}${\small Laboratoire associ\'e au CNRS--URA-D0063.}\\
April 1995 \end{flushleft} \thispagestyle{empty}
\end{titlepage}}
\newcommand{\C}{{\cal C}}
\newcommand{\B}{{\cal B}}
\newcommand{\NIJ}{{\cal N}_{IJ}}
\newcommand{\N}{{\cal N}}
\newcommand{\dal}{\raisebox{0.085cm}
{\fbox{\rule{0cm}{0.07cm}\,}}}
\newcommand{\dt}{\partial_{\langle T\rangle}}
\newcommand{\dtbar}{\partial_{\langle\fracline{T}\rangle}}
\newcommand{\al}{\alpha^{\prime}}
\newcommand{\mst}{M_{\scriptscriptstyle \!S}}
\newcommand{\mpl}{M_{\scriptscriptstyle \!P}}
\newcommand{\dv}{\int{\rm d}^4x\sqrt{g}}
\newcommand{\lv}{\left\langle}
\newcommand{\rv}{\right\rangle}
\newcommand{\ph}{\varphi}
\newcommand{\sbar}{\,\fracline{\! S}}
\newcommand{\xbar}{\,\fracline{\! X}}
\newcommand{\barz}{\,\fracline{\! Z}}
\newcommand{\zbar}{\bar{z}}
\newcommand{\dbar}{\,\fracline{\!\partial}}
\newcommand{\tbar}{\fracline{T}}
\newcommand{\psibar}{\fracline{\Psi}}
\newcommand{\ybar}{\fracline{Y}}
\newcommand{\z}{\zeta}
\newcommand{\zb}{\bar{\zeta}}
\newcommand{\phb}{\fracline{\varphi}}
\newcommand{\cm}{Commun.\ Math.\ Phys.~}
\newcommand{\pr}{Phys.\ Rev.\ D~}
\newcommand{\pl}{Phys.\ Lett.\ B~}
\newcommand{\np}{Nucl.\ Phys.\ B~}
\newcommand{\e}{{\rm e}}
\newcommand{\gsi}{\,\raisebox{-0.13cm}{$\stackrel{\textstyle
>}{\textstyle\sim}$}\,}
\newcommand{\lsi}{\,\raisebox{-0.13cm}{$\stackrel{\textstyle
<}{\textstyle\sim}$}\,}
\date{}
\firstpage{3117}{hep-th/9504143}
{\large\bf Dynamical Supersymmetry
Breaking and the Linear Multiplet$^\star$}
{P. Bin\'etruy,$^{\,a}$ M.K. Gaillard$^{\,b}$ $\,$and$\,$
T.R. Taylor$^{\,c}$}
{\normalsize\sl
$^a$Laboratoire de Physique Th\'eorique et Hautes Energies,$^{\dagger}$
Universit\'e Paris-Sud,\\[-5mm]
\normalsize\sl F-91405 Orsay, France\\
\normalsize\sl $^b$Department of Physics and Theory Group, Lawrence
Berkeley Laboratory, \\[-5mm]
\normalsize\sl University  of California, Berkeley 94720, U.S.A.\\
\normalsize\sl $^c$Department of Physics, Northeastern
University, Boston, MA 02115, U.S.A.}
{We analyze gaugino condensation in the presence of a dilaton and an
antisymmetric tensor field, with couplings reminiscent of string
theories. The degrees of freedom relevant to a supersymmetric description of
the effective theory below the scale of condensation
are discussed in this context.}

\setcounter{section}{0}
\section{Introduction}
The dilaton and antisymmetric tensor fields that are found among
massless string modes are believed to play an important role
in the dynamics of supersymmetry breaking.
Together with their supersymmetric partners, they form a linear
supermultiplet that is described
by a vector superfield $L$ satisfying the constraint
$\bar{D}^2L={D}^2L=0$, in rigid supersymmetry notation.
Thus, the vector component ${\cal B}^m$ satisfies
\begin{equation}
\partial_m{\cal B}^m=0\ , \label{eq:b}
\end{equation}
and is therefore constrained to describe the field strength of a
Kalb-Ramond field (two-index antisymmetric tensor field):
${\cal B}^m=\epsilon^{mnpq}\partial_nb_{pq}$.

The generation of masses for the dilaton and the two-form
presents one of the most important problems
faced by superstring phenomenology. It must have a
non-perturbative nature since the linear multiplet remains
massless to all orders in the string loop expansion.
This problem has been addressed before
in the context of dynamical supersymmetry
breaking induced by gaugino condensation. The standard strategy
is to first perform a duality transformation replacing the linear
superfield by a chiral one, and thus replacing the antisymmetric
tensor by a pseudoscalar axion. A superpotential for the chiral
superfield can then be derived in the framework of effective field
theory. However the analysis of the induced potential is plagued
by serious difficulties such as the absence of a stable vacuum
and a vanishing axion mass.

Although the equivalence between linear and chiral
multiplets is certainly valid for on-shell amplitudes,
it is not clear that it holds in off-shell
quantities like non-perturbative effective potentials.
It is therefore very important to consider gaugino condensation
with the dilaton and antisymmetric tensor, without recurring
to any duality transformation. In particular, it has been known
for quite a long time that a non-zero mass for the antisymmetric
tensor can be generated by couplings to a vector or to a
three-form (three-index antisymmetric tensor field).
In this paper, we show that in the effective theory
of gaugino condensation,
these extra degrees of freedom
are supplied by gauge-singlet composite fields. The effective
theory is written in terms of a single vector multiplet
which contains these composite fields
in addition to the degrees of freedom contained in the linear multiplet.

The antisymmetric tensor field plays a very important role in the anomaly
cancellation of heterotic superstring theory. As a consequence,
in the Lagrangian describing the massless string modes, the
linear multiplet always appears in the gauge-invariant combination
$L-\Omega$, where $\Omega$ is the Chern-Simons (real) superfield.
The Yang-Mills part of $\Omega$ is defined by
$\bar{D}^2\Omega= -k W^{\alpha}W_{\alpha}$ with $k$ a
normalisation factor, so that
\begin{equation}
\bar{D}^2(L-\Omega)= k W^{\alpha}W_{\alpha}\ , \label{eq:ww}
\end{equation}
where $W^{\alpha}$ is the usual gauge field-strength chiral superfield.

In the effective theory  of gaugino condensation, we
replace the superfield $k W^\alpha W_\alpha$ by an effective
chiral superfield
$U$ describing the condensate degrees of freedom \cite{VY}. As a consequence of
the Bianchi identities
in the Yang-Mills sector, this superfield must derive from a real
``prepotential'' $V$, that is
\begin{equation}
\bar{D}^2V=U\  . \label{eq:u}
\end{equation}
By comparison with eq.(\ref{eq:ww}), we see that the single
vector superfield $V$ describes the ``elementary'' degrees of freedom
of the linear multiplet together with the ``composite'' degrees of
freedom of the gauge singlet bound states.
This vector supermultiplet
will be the central focus of the following analysis.

The essence of the non-perturbative analysis is contained in a superpotential
term for $U= \bar{D}^2 V $ that is dictated by the
anomaly structure of the underlying theory \cite{VY}.
The full effective action is
then completed by replacing  the gauge invariant combination
$L- \Omega$ by $V$ in the original action.

The next section is an illustration of these ideas in the context of
global supersymmetry. The problem of the determination of the form of
the kinetic terms for the effective degrees of freedom cannot be solved
at this level using only symmetry arguments. We therefore consider in
section 3 the context of superstring theories where higher derivative
terms in the action of the underlying theory may provide the right
kinetic terms for the dynamical
degrees of freedom.

\section{The global supersymmetry case.}

In this section, we start with  global supersymmetry  and a
toy model that reproduces some of the basic properties of the more
realistic supergravity models to be considered in what follows.  We show
in particular that gaugino condensation in presence of a dilaton field
$\cal C$ and an antisymmetric tensor $b_{mn}$ induces a non-trivial
potential for the dilaton field, in agreement with the results obtained
previously in the dual formulation.

Let us consider a general vector supermultiplet $V$. We will write
its  component field expansion in superspace as
\begin{eqnarray}
V &=& {\cal C} + i \theta \chi - i \bar{\theta} \bar{\chi} +
\theta^2 h + \bar{\theta}^2 h^* + \theta \sigma^m
\bar{\theta} {\cal B}_m \nonumber\\
&+& i\theta^2 \bar{\theta}(\bar{\eta} + {i \over 2} \bar{\sigma}^m
\partial_m\chi) -i\bar{\theta}^2 \theta ( \eta + {i \over 2}
\sigma^m \partial_m \bar{\chi}) + {1\over2}\theta^2 \bar{\theta}^2
(D + {1 \over 2} \Box {\cal C}) \label{eq:expansion}
\end{eqnarray}
The standard way to proceed from there is to impose a constraint in
superspace, the so-called linear multiplet constraint $D^2 V =
\bar{D}^2 V = 0$ which yields in component form
\begin{equation}
\partial^m {\cal B}_m = 0, \label{eq:constraint}
\end{equation}
together with the vanishing of all auxiliary fields.
Eq.(\ref{eq:constraint}) is the Bianchi identity which ensures
that ${\cal B}_m$ is  dual to the field strength of an antisymmetric tensor:
\begin{equation}
{\cal B}_m = \epsilon_{mnpq} \partial^n b^{pq}. \label{eq:fstrength}
\end{equation}
We will instead obtain the Bianchi identity (\ref{eq:constraint})
as an equation of motion.

Let us indeed consider the following Lagrangian for $V$
\begin{equation}
{\cal L} = \int d^4\theta K(V) + \left( \int d^2\theta W(\bar{D}^2 V)
+ {\rm h.c.} \right), \label{eq:lag1}
\end{equation}
where $K$ is a real function of $V$ whereas $W$ is an analytic
function of the chiral superfield $\bar{D}^2V$, which will turn out
to be our prepotential field.

Since we are mainly interested in applying our ideas to the string
dilaton-axion-dilatino system, we will make the following choices
\begin{equation}
K(V) = \log V, \;\;\; W(U) = - {b \over 4} \; U \log U ,\label{eq:KW}
\end{equation}
where $U=\bar{D}^2 V$. The form of the superpotential is dictated by
the anomaly structure of the underlying theory and $b$ is
proportional to the one-loop beta function coefficient.

The Lagrangian (\ref{eq:lag1}) can in fact be obtained through a duality
transformation as follows. Consider the alternative Lagrangian:
\begin{equation}
{\cal L} = \int d^4\theta \left( K(V) + (V-\Xi)(S+\bar{S})\right)
+ \left( \int d^2\theta W(\Delta)
+ {\rm h.c.} \right), \label{eq:lag1'}
\end{equation}
where $S$ and $\Delta$ are chiral superfields which are further
constrained by requiring that $\Delta$ derives from the {\em real}
prepotential $\Xi$; in other words,
\begin{eqnarray}
\Delta &=& \bar{D}^2 \Xi, \nonumber \\
\bar{\Delta} &=& D^2 \Xi.
\end{eqnarray}
We will see later that such a chiral superfield naturally arises when
one tries to describe a 3-form supermultiplet in supersymmetry or
supergravity. For the time being, in the context of gaugino
condensation, $\Delta$ can be thought as a relic Tr$W^\alpha W_\alpha$
and $\Xi$ as a relic Chern-Simons superfield. The two are obviously
connected. This identification
becomes more transparent if one minimizes with respect to $V$, in which
case one finds simply
\begin{equation}
V = - {1 \over S + \bar{S}}
\end{equation}
and the Lagrangian (\ref{eq:lag1'}) can be written as
\begin{equation}
{\cal L} = \int d^4\theta \left[-\log(S+\bar{S})\right]
+ \left( \int d^2\theta (4S\Delta + W(\Delta)) + {\rm h.c.} \right).
\label{eq:lag1''}
\end{equation}
Note that we have not included a kinetic term for $\Delta$. In other
words, we do not consider the dynamics of composite fields in this
first approach. We will  return below to this question.

One can alternatively minimize the Lagrangian (\ref{eq:lag1'}) with
respect to $S$, which yields
\begin{eqnarray}
\bar{D}^2V =& \bar{D}^2\Xi =& \Delta, \nonumber \\
D^2V =& D^2 \Xi =& \bar{\Delta},
\end{eqnarray}
in which case one recovers the original Lagrangian (\ref{eq:lag1}).

Let us now write the Lagrangian (\ref{eq:lag1}) in terms of the
component fields (we will disregard here the fermion fields):
\begin{eqnarray}
{\cal L}_{bos} &=& -{1 \over 4 {\cal C}^2} \partial^m {\cal C}
\partial_m {\cal C} + {1 \over 4 {\cal C}^2} {\cal B}^m {\cal B}_m
+ \left( {1 \over 2 {\cal C}}
+ b (1+ \log |4h|) \right) (D + \Box {\cal C})\nonumber \\
&&- {1 \over {\cal C}^2} |h|^2 + i{b \over 2} \log{h^* \over h}
\partial^m {\cal B}_m \label{eq:lag1comp}
\end{eqnarray}
Writing $h \equiv \rho e^{i\omega}$, we obtain by minimizing with
respect to the phase $\omega$ precisely the Bianchi identity
\begin{equation}
\partial^m {\cal B}_m =0 \label{eq:Bid}
\end{equation}
which ensures that ${\cal B}_m$ is dual to the field strength of an
antisymmetric tensor according to (\ref{eq:fstrength}). The term
${\cal B}^m{\cal B}_m$ in the Lagrangian is therefore a kinetic term for
this tensor field.

Minimizing with respect to $D$ yields
\begin{equation}
\rho = {1\over 4e} e^{-{1 \over 2 b {\cal C}}}.
\end{equation}
The Lagrangian now reads
\begin{eqnarray}
{\cal L} &=& -{1 \over 4 {\cal C}^2} \partial^m {\cal C} \partial_m
{\cal C} + {1 \over 4 {\cal C}^2} {\cal B}^m{\cal B}_m-V({\cal C}) \nonumber \\
V({\cal C}) &=& {1 \over 16 e^2 {\cal C}^2} e^{-{1 \over b{\cal C}}}
\label{eq:pot1} \end{eqnarray}
Finally, minimizing (\ref{eq:lag1comp}) with respect to $\rho$ yields
\begin{equation}
D = {1 \over 8 b e^2 {\cal C}^2} e^{-{1 \over b {\cal C}}}\ .
\end{equation}
The fact that we are left with an auxiliary field
$D$ is another departure from the standard linear multiplet treatment,
which is known not to include any auxiliary field.

Let us note that our treatment is very dependent on the form
(\ref{eq:KW}) of the superpotential $W$. It can easily be shown that
only such a form - up to a constant term - allows one to recover the
Bianchi identity (\ref{eq:constraint}) through the minimization
condition on the phase of the $h$ field.

In a more realistic model which includes kinetic terms for the
composite degrees of freedom, the simple Bianchi identity
(\ref{eq:Bid}) no longer holds. It is replaced by
\begin{equation}
\partial^m {\cal B}_m = {1 \over 8} \; {}^* \Phi \label{eq:Bid'}
\end{equation}
where ${}^* \Phi$ is related to the field strength of a rank-3
antisymmetric tensor field $\Gamma_{npq}$, remnant of the Chern-Simons
form in the effective theory:
\begin{equation}
{}^* \Phi = {1 \over 3!} \epsilon_{mnpq} \partial^m \Gamma^{npq}.
\label{eq:Phi}
\end{equation}
Indeed, in this context, the fact that the scalar field $U$ derives
from a real prepotential $V$ through (\ref{eq:u}) allows us to interpret
its degrees of freedom as those of a 3-form supermultiplet
\cite{Gates,GG,BGGP}. Such a supermultiplet is precisely constructed with the
help
of a scalar field $U$ which satisfies the constraint\footnote{Such
a supermultiplet is constructed from a super
3-form gauge potential $\Gamma_{ABC}$ ($A, B, C$ being vector or spinor
indices). Its field strength $\Phi_{ABCD}$ is constrained by
$\Phi_{\alpha \beta \gamma A} = 0$ ($\alpha, \beta, \gamma$ dotted or
undotted spinor indices). The analysis of the Bianchi identity $d\Phi =
0$ shows that all the other components are expressed in terms of a
single chiral superfield $U$ defined by:
${\Phi^{\dot{\alpha}\dot{\beta}}}_{ab} = 16\; (\bar \sigma _{ab}
\epsilon)^{\dot{\alpha}\dot{\beta}} U$. For example, $\Phi_{abcd} = -
\epsilon_{abcd} {}^* \Phi$ with $2 i {}^* \Phi = D^2 U - \bar D^2 \bar
U$.}
\begin{equation}
D^2 U - {\bar D}^2 {\bar U} = 2i \; {}^* \Phi, \label{eq:3form}
\end{equation}
where ${}^* \Phi$ is the gauge invariant field strength of the 3-form
as in (\ref{eq:Phi}). One can show that the relation (\ref{eq:3form})
holds only for a chiral superfield $U$ deriving from a real
prepotential such as in $U = {\bar D}^2 V$.

The solution to (\ref{eq:Bid'}) is
\begin{equation}
{\cal B}_m = {1 \over 3!} \epsilon_{mnpq} \left( {1 \over 8}
\Gamma^{npq} + \partial^n b^{pq} + \partial^q b^{np}
+ \partial^p b^{qn} \right),
\end{equation}
where $b^{pq}$ is a 2-form. This 2-form can be gauged away by
performing a gauge transformation on the 3-form:
\begin{equation}
\Gamma^{npq}\rightarrow\Gamma^{npq}+ \partial^{[n} \Lambda^{pq]}\ .
\label{gauge} \end{equation}
With this choice of gauge, the vector ${\cal B}_m$
can be interpreted
as a field dual to a 3-form,
\begin{equation}
{\cal B}_m =  {1 \over 8 \cdot 3!} \;
\epsilon_{mnpq} \Gamma^{npq}.\label{gamma}
\end{equation}

In order to be more specific, we need to write the
kinetic term for the composite degrees of freedom. The problem is that
symmetry considerations do not restrict this term in any significant way
and, as long as we do not place ourselves in the context of a given
theory (see next section), we have little information about such a
term. We will therefore choose the generic form
\begin{equation}
{\cal L}_{kin} = \alpha \int d^4\theta \; V^n ({\bar D}^2 V D^2 V)^p.
\end{equation}
The full Lagrangian now reads in component form:
\begin{eqnarray}
{\cal L} &=& - f(\rho, \C) \; \partial^m \C \partial_m \C  - \alpha \C^n
(16\rho^2)^p ({\partial^m \rho \partial_m \rho \over \rho^2} +
\partial^m \omega \partial_m \omega) \cr
& & + 4 \alpha p^2 \C^n(16\rho^2)^{p-1} (\partial^m \B_m)^2
+ f(\rho,\C) \; \B^m \B_m  - [b + \alpha np \;
\C^{n-1} (16\rho^2)^p] \B^m \partial_m \omega \cr
& & - 4 f(\rho,\C) \; \rho^2
+ 4 \alpha p^2 \C^n (16\rho^2)^p (D+ \Box \C)^2 \cr
& &
+ \left[ {1\over 2 \C}(1 + \alpha n (2p+1) \C^n (16\rho^2)^p) + b
(1 + \log (4\rho)) \right] (D + \Box \C)  \label{eq:fulag}
\end{eqnarray}
where
\begin{equation}
f(\rho,\C) = {1 \over 4 \C^2} [1 - \alpha \; n(n-1) \C^n (16\rho^2)^p].
\end{equation}
In eq.(\ref{eq:fulag}), the auxiliary field $D$ can be
eliminated in a simple way by using its equations of motion.

In eq.(\ref{eq:fulag}), the vector field degrees of freedom are
represented by a 3-form, as in eq.(\ref{gamma}).
The field equation obtained by varying $\Gamma_{npq}$
can be expressed in terms of the scalar field:
\begin{equation}
a \equiv {1 \over 2} \alpha p^2 \C^n (16\rho^2)^{p-1} \; {}^* \Phi\ .
\end{equation}
It reads
\begin{equation}
\partial^k \left( {2\C^2 \left[2\partial_k a + (b + \alpha
np\; \C^{n-1}(16\rho^2)^p ) \; \partial_k \omega \right] \over
1 - \alpha n(n-1)\C^n (16\rho^2)^p}\right) = {a \over 4
\alpha p^2 \C^n (16\rho^2)^{p-1}}
\end{equation}
This has the form of an equation of motion for the scalar $a$,
mixed with the phase $\omega$ at the
level of the kinetic terms, {\it i.e.},
\begin{equation}
{\cal L}_{kin} = - {\C^2  \over1 - \alpha n(n-1)\C^n (16\rho^2)^p}
\left( 2\partial^m a +  (b + \alpha np\; \C^{n-1}(16\rho^2)^p
)\partial^m \omega \right)^2,
\end{equation}
and whose non-derivative
interactions are described by the potential \begin{equation}
v(a) = {a^2 \over  4 \alpha p^2
\C^n(16\rho^2)^{p-1}}.\label{eq:massterm} \end{equation}
Note that $a$ represents a gauge-invariant degree of freedom
from the point of view of the 3-form gauge transformation
(\ref{gauge}). We therefore interpret $a$ as an axion-like degree of
freedom associated with the linear multiplet of the fundamental
theory. The potential term (\ref{eq:massterm}) thus represents a mass
generation for the corresponding antisymmetric tensor field. And the
equation of motion for $\Gamma_{npq}$ is dual to the equation of
motion of a massive axion.

The Lagrangian (\ref{eq:fulag}) therefore describes the interaction of
two -- dilatonic and axionic -- fundamental degrees of freedom $\C$ and
$a \sim \partial^m {\cal B}_m$ and two -- dilatonic and axionic --
effective degrees of freedom $h$ and $\omega$, describing gaugino
condensates. The fundamental axionic degree of freedom can also be
interpreted as an effective gauge condensate $\lv F^{mn} {\tilde F}_{mn}\rv$.

It is quite interesting to note that the couplings in (\ref{eq:fulag})
responsible for the mass generation of the axionic field $a$ were
proposed earlier \cite{AT,ATT} on the basis of a parallel with the
Schwinger model in two dimensions. This connection sheds some light on
the reason why no massless particle appears in the physical spectrum of
the theory.

The minimum of the full scalar potential occurs at $a=0$. Further minimization
with respect to $\rho$ and ${\cal C}$ yields
\begin{eqnarray}
(1 + \alpha n (2p+1) \C^n (16\rho^2)^p) + 2 b \; \C
(1 + \log (4\rho)) &=& 0, \cr
f(\rho,\C) = {1 \over 4 \C^2} [1 - \alpha \; n(n-1) \C^n (16\rho^2)^p]
&=& 0.
\end{eqnarray}
The first equation implies zero vacuum expectation value
of the auxiliary field $D$, while the second leads to a
vanishing normalisation factor of the kinetic energy term for the
$\C$ field, as can be seen from (\ref{eq:fulag}). The minimization
procedure thus leads in this simple model to a singular point of the
field configuration. We now turn to more realistic configurations where
the form of the kinetic terms can be inferred from the structure of an
underlying fundamental theory.

\section{Local Supersymmetry}
\noindent {\it 3.1. Minimal Terms}

We will now consider a locally supersymmetric model for the
prepotential $V$ describing the linear multiplet together with the
composite degrees of freedom.
The Lagrangian will be constructed by using the formalism
of superconformal supergravity, with  $N{=}1$
Poincar\'e gauge fixing constraints imposed on the chiral compensator
\cite{KU}.
Recall that in this formalism, the multiplets are characterized
by their dimensions and $U(1)$ charges: real vector superfields
are neutral whereas the charges of chiral superfields
are equal to their dimensions.  The chiral compensator
$\Sigma$ has dimension 1 while the real linear multiplet,
hence also the prepotential $V$, has dimension 2.
The fact that $V$ carries a non-zero dimension, which is a consequence
of its gravitational origin, is very useful for discussing
various limiting cases of the effective theory.
In general,
a locally supersymmetric action can be constructed from an F-component
of a chiral superfield of dimension 3 or from
a D-component of a real vector multiplet of dimension 2.

Once the gauge-invariant combination $L-\Omega$ is replaced
by the vector  superfield $V$, the tree-level kinetic
terms of the dilaton become
\begin{equation}     \label{tree}\left.
-(\Sigma\bar{\Sigma})^{3/2}\left(\frac{V}{2}\right)^{-1
/2}\right|_{\makebox D}
=\left.\sqrt{2}\,{\cal P}[(\Sigma\bar{\Sigma})^{3/2}
V^{-1/2}]\right|_{\makebox F} +\makebox{h.c.,}
\end{equation}
where ${\cal P}$ is the chiral projection operator -- a
generalization of $-\bar{D}^2/2$ to the local case.\footnote{Hence a
factor $-$2 rescaling on the global superspace measure $d^2\theta$ is
necessary in order to recover the conventions of
ref.\cite{KU} used in this section.}
It has dimension 1 and charge 3.

As already mentioned,
the form of the non-perturbative superpotential is dictated
by the anomaly structure of the underlying gauge theory.
It gives rise to the interaction term
\begin{equation}\label{super}\left.
-b{\cal P} [V] \ln (-2{\cal P} [V]/\Sigma^3)\right|_{\makebox F}
 +\makebox{h.c.}
\end{equation}

In the following discussion, we will be interested mostly
in the bosonic terms in the Lagrangian,
in particular in the scalar potential.
The bosonic components of $V$ are, as in the global case, the dilaton
$\cal C$, the vector ${\cal B}^m$ and two auxiliary fields,
$h$ as an F-component and $D$ as a D-component.
We also define
\begin{equation}\label{pi}
\Pi\equiv\sqrt{2}\,{\cal P}
[(\Sigma\bar{\Sigma})^{3/2}V^{-1/2}]\, ,
\end{equation}
a chiral superfield of dimension 3.

The most convenient gauge choice for the
scalar component $\sigma$ of the compensator is
\begin{equation}\label{sigma}
\sigma^3=\sqrt{{\cal C}/2}
\end{equation}
The bosonic part of the Lagrangian
\begin{equation}\label{l}
{\cal L} ~=~ \Pi|_{\makebox F} \left. -b{\cal P} [V] \ln
(-2{\cal P} [V]/\Sigma^3)\right|_{\makebox F}
+\makebox{h.c.}
\end{equation}
is given in this gauge by
\begin{eqnarray} \label{bos}
{\cal L}_{\makebox{\scriptsize bos}} &=&
-\frac{1}{2}R-\frac{1}{4{\cal C}^2}\partial^m {\cal C}
\partial_m {\cal C}
+\frac{1}{4{\cal C}^2} {\cal B}^m {\cal B}_m
+\frac{1}{8}(3A^m-\frac{{\cal B}^m}{\cal C})^2\nonumber\\[3mm] & &
+\left[ \frac{1}{2{\cal C}}
+ \frac{b}{2} \left( 2+\log \frac{8|h|^2}{\C} \right) \right]
(D+\Box\C+\frac{\C}{3}R) \nonumber\\[3mm] & &
-\frac{|h|^2}{{\cal C}^2}-\frac{|\tilde{h}|^2}{2}
- b (\tilde{h}h^*+\tilde{h}^*h+
\frac{2|h|^2}{\C}) +  i \frac{b}{2} \log \frac{h^*}{h} \partial_m{\cal
B}^m \, ,
\end{eqnarray}
where $R$ is the Ricci scalar and $A^m$ is the superconformal
$U(1)$ gauge field; $A^m$ is an auxiliary field which
becomes $A^m=\frac{{\cal B}^m}{3\C}$ upon using its equations of motion.
Finally, $\tilde{h}$ is a mixture of the auxiliary component
$h_{\Sigma}$ of the chiral compensator and of $h$.

Actually comparison  with the global case, eq.(\ref{eq:lag1comp}), shows
that the only major modification comes from mixed contributions
between this auxiliary field and $h$. The equation of motion for $D$,
on the other hand, fixes the standard exponential behavior for $h$ in
terms of the dilaton $\C$.
Solving for $D$ and $\tilde{h}$ one obtains the following dilaton
potential:
\begin{equation}
V(\C) = {1 \over 8e^2} \left( {1 \over \C} + 2 b - 2 b^2 \C
\right) \; e^{-{1 \over b \C }},
\end{equation}
very similar to the global case.

As before, the question of central interest is the origin of the
kinetic terms for the effective degrees of freedom, which obviously do
not appear in the simple terms that we have considered so far, eq.
(\ref{l}). We now turn to this problem.

\vskip .8cm
\noindent {\it 3.2. Kinetic terms for the composite degrees of freedom.}

The central object in our discussion will be the chiral superfield
$\Pi$ introduced above in eq.(\ref{pi}). If we work in string units,
then $V$ is of order $g^0$ and $(\Sigma \bar \Sigma )^{3/2}$ is of
order $g^{-2}$, where $g$ is the string coupling constant. Thus $\Pi$
is of order $g^{-2}$ as well and the term $\Pi|_F$ which we used in
the preceding subsection appears at the string tree level.

Kinetic terms for the composite fields may be expected at the one-loop
level. Indeed, one may construct from $\Pi$ the real superfield $\Pi
\bar \Pi V / (\Sigma \bar \Sigma)^3$ of conformal weight 2, whose
D-component therefore is a natural candidate for a locally
supersymmetric action term. Such  a term is expected to appear at the
string one loop-level (order $g^0$ when working in string units). As we
will see momentarily, it naturally provides a kinetic term for the
gauge and gaugino condensates, when interpreted in terms of the degrees
of freedom present in the effective theory.

Moreover, higher weight generalized superpotential interactions \cite{Ant} are
expected on the same ground. They are again constructed from the
superfield $\Pi$, the chiral projection of the vector superfield \cite{KU}.
Indeed, the chiral superfield $\Sigma^3 (\Pi / \Sigma^3)^n$ has
conformal weight 3 and therefore leads, through its F-term, to a
superpotential interaction. In string units, it is of order $g^{-2}$
and therefore all the corresponding interactions appear at the string
tree level \cite{Ant}. We will express them through a general
function $\Sigma^3 F[\Pi / \Sigma^3]$. The linear term in the
expansion of $F$ ($n=1$) provides just the term $\Pi|_F$ that
was included in the action of the preceding subsection.

To recapitulate, we start with the more complete action:
\begin{equation}
{\cal L} = \left. \alpha {V \over (\Sigma \bar \Sigma)^3} \Pi \bar \Pi
\right|_{\makebox{D}}-\left[\left. \left( \Sigma^3 F\left({\Pi \over \Sigma^3}
\right)
+ b{\cal P} [V] \ln (-2{\cal P} [V]/\Sigma^3)\right)\right|_{\makebox F}
+{\rm h.c.}
\right]
\end{equation}
where $\alpha$ is some normalisation constant that we will take to be
positive. We believe that this action represents the terms computable
from string interactions that, interpreted in terms of the dynamical
degrees of freedom, describe the  effective theory below the
condensation scale.

We could follow the same procedure as in the global case, that is,
we could write the corresponding equations of motion for the different fields
(in particular the 3-form and the 2-form) and infer from them the
expression for the potential energy in terms of the corresponding
degrees of freedom. Being only interested here in this potential
energy, we will depart somewhat from our original orientation and
perform what amounts to a duality transformation in order to shorten the
derivation.

Indeed, one can show that the same theory can be derived from
the following Lagrangian:
\begin{eqnarray}
{\cal L} &=& \left.\left[ \alpha {V \over (\Sigma \bar \Sigma)^3} \Pi
\bar \Pi - \frac{(S + \bar S)}{2} V - (X + \bar X) \left(\frac{V}{2}
\right)^{-1/2} (\Sigma \bar
\Sigma)^{3/2} \right] \right|_{\makebox{D}} \nonumber\\[3mm]
&+& \left( \left. \left[- \Sigma^3
F\left({\Pi \over \Sigma^3} \right) +X \Pi + \frac{S U}{2} +
\frac{b}{2} U \log
{U \over
\Sigma^3}\right] \right|_{\makebox{F}} + {\rm h.c.} \right), \label{eq:Lcomp}
\end{eqnarray}
where $S$ and $X$ are chiral superfields. Minimization with respect to
$S$  ensures the constraint
$U=-2{\cal P} [V]$,
whereas minimization with
respect to $X$ ensures that $\Pi$ is given as in (\ref{pi}). Thus,
$\Pi$ in (\ref{eq:Lcomp}) is to be taken as an independent chiral
superfield.

Now, solving for $V$ and $U$ yields:
\begin{equation}
{\cal L} = - \left.{3 \over 2} (\Sigma \bar \Sigma) e^{-{1 \over 3} K}
\right|_{\makebox{D}} + \left( \left. \Sigma^3 W
\right|_{\makebox{F}} + \;{\rm h.c.}\right),
\end{equation}
where the K\"ahler potential is given, up to a constant piece, by
\begin{equation}
K = - \log (S + \bar S - \alpha \Pi \bar \Pi) - 2 \log (X + \bar X)
\end{equation}
and the superpotential by
\begin{equation}
W =  X \Pi -F(\Pi) - {b \over 2e} e^{-S/b}; \label{eq:W}
\end{equation}
The field $\Pi/\Sigma^3$ has been redefined as $\Pi$.

The presence of new propagating superfields $X$ and $\Pi$ can be
understood in the following way. The higher weight interactions
\cite{Ant} give rise to kinetic terms for fields that appear to be
auxiliary at low energies, like the $h$ component of $V$. At high
energies, these fields form full physical supermultiplets
corresponding to superstring excitations. $X$ is an example of such a
superfield, with its mass equal to the superstring mass scale. The
superfield $\Pi$ contains the composite degrees of freedom of the
gaugino bound state and its supersymmetric partners.\footnote{
This can also be viewed from a description \`a la
Nambu-Jona-Lasinio. As can be seen from (\ref{eq:Lcomp}),
the field $X$ can be understood as an ``auxiliary'' field whose
equation of motion ensures that $\Pi$ describes the gaugino
condensate degrees of freedom. Similar couplings appear in a
supersymmetric version of the Nambu-Jona-Lasinio scenario\cite{MC}. We wish
to thank Yi-Yen Wu for pointing this out to us.}

With the theory now formulated in the standard supergravity framework,
it is straightforward to obtain the scalar potential.
As one can easily check, it suffers from an instability as $S + \bar S
- \alpha \Pi \bar \Pi \rightarrow 0^+$.
However this is not a genuine problem of this approach and it can
be cured in many ways. As an example, we
include in the model a modulus $T$ which appears only in the K\"ahler
potential. We therefore compute the scalar potential with
\begin{equation}
K = - \log (S + \bar S - \alpha \Pi \bar \Pi) - 2 \log (X + \bar X)
 - 3 \log (T + \bar T)
\end{equation}
and the superpotential of (\ref{eq:W}). It reads
\begin{eqnarray}
V &=& - {1 \over 2 Y}|\Pi|^2 + {1 \over Y (X + \bar X)} (\Pi \bar F +
\bar \Pi F) \cr
& & + {1 \over Y (X + \bar X)^2} \left( 3 |F-X\Pi|^2 +{Y \over \alpha}
|F'-X|^2 + \left[ Y (S + \bar S + 2b) + 3b^2 \right] e^{-{S + \bar
S \over b}-2} \right. \cr
& & \;\;\;\;\;\;\;\;\;\;\;\;\;\;\;\;\;\;\;\;\;\;\; \left. - \left[ e^{-{S
\over b}-1}\left(  (3b + Y) \bar F - Y \bar \Pi {\bar F}' + b \bar \Pi
(X - 2 \bar X)\right) + {\rm h.c.} \right] \right).
\end{eqnarray}
where $Y=S + \bar S - \alpha \Pi \bar \Pi$ and $F'  = dF/d\Pi$.

In the simple case where the function $F$ is linear in $\Pi$, one can
minimize with respect to all the fields but $s \equiv S + \bar
S$ and $\rho \equiv |\Pi|$. The potential then has a smooth monotonic
behaviour with respect to $s$ and $\rho$, going to $0^+$ as any of
these fields tends to infinity. There is no non-trivial minimum.
On the other hand, there is no problem with generating the Im$\,S$
axion mass: the Peccei-Quinn symmetry
$S\rightarrow S+ia$
is broken by the superpotential
terms involving $\Pi$ and the massive string superfield $X$.
We reserve the study of more complete and realistic models to further work.

\section{Concluding remarks}

The vector supermultiplet formalism developed in this work allows
a very natural effective action description
of the coupled systems of gauge fields and
linear supermultiplets, as present in the heterotic superstring
compactifications.
In particular, the vector component of this multiplet contains
a three-form field that can
describe either the field-strength of a massless Kalb-Ramond field
or a massive axion, depending on the details of non-perturbative dynamics.
Furthermore, the formalism allows straightforward incorporation
of the higher-weight generalized superpotential interactions.
In the equivalent dual description, the axion mass generation
can be understood as an effect of non-perturbative superpotential
terms that violate a Peccei-Quinn-like symmetry.

While (slowly) completing this article, we
received a paper by C. Burgess, J.P. Deren\-din\-ger, F. Quevedo and M.
Quir\'os \cite{BDQQ} where ideas similar to the ones presented here are
developed, although with seemingly different motivations and only in
the context of global supersymmetry. An earlier and somewhat
different approach has been followed by Gaida and L\"ust \cite{GL}.

\vskip 1cm
\noindent
{\bf Acknowledgments}

P.B.\ and T.R.T.\  would like to thank Theory Divisions at LBL
and at CERN for their hospitality. P.B.\ also wishes to thank G.
Girardi and R. Grimm for enlightening discussions on the 3-form
supermultiplet.

\end{document}